\documentclass[a4paper,11pt]{article}
\usepackage{jheppub} % for details on the use of the package, please see the JINST-author-manual
\usepackage{lineno}
\newcommand{\bea}{\begin{eqnarray}}
\newcommand{\eea}{\end{eqnarray}}
\newcommand{\nnr}{\nonumber \\}

\title{\boldmath Does SO(2,2) symmetry lead to an additional U(1) symmetry in 5-dimensional Einstein-scalar theories?}

\author{Saeedeh Sadeghian}
\affiliation{Department of Theoretical Physics, Faculty of Science, University of Mazandaran,\\
P.O. Box 47416-95447, Babolsar, Iran}

\emailAdd{s.sadeghian@umz.ac.ir}

\abstract{Solving Einstein equations including the matter fields in dimensions higher than four is computationally challenging. However, the existence of symmetries facilitates these calculations. Among these symmetries, we can mention the SO(2,2) symmetry that arises in the near-horizon geometries of extremal black holes with vanishing horizon area in four dimensions \cite{EVH/CFT} and higher dimensions \cite{Sadeghian:2015}. Solutions of Einstein theories including scalar and Maxwell fields with SO(2,2) symmetry have been classified in four, five, and higher dimensions \cite{Sadeghian:2014}. For the Einstein-Maxwell-Lambda theory, a theorem states that SO(2,2) symmetry leads to an additional U(1) symmetry and therefore unique axisymmetric solutions. Nevertheless, in the presence of a scalar field, this symmetry group is not large enough in five dimensions to guarantee solution uniqueness, and the additional U(1) symmetry remains a conjecture in \cite{Sadeghian:2014}. Here, we study the Einstein equations of theories with a five-dimensional scalar field using the Lie symmetry method for solving partial differential equations. We challenge the conjecture in \cite{Sadeghian:2014} and find a solution that depends on both $\theta$ and $\psi$ coordinates as a counterexample to that conjecture.}

\begin{document}
\maketitle
\flushbottom

\section{Introduction}
%@@@@@@@@@@@@@@@@@@@@@@@@@@@@@@@
Symmetries can significantly simplify the solution of a problem. Symmetries are usually discussed at the level of action or at the level of solutions to the equations of motion. However, symmetries have been studied at the level of the equations of motion. This helps in solving the differential equations that describe motion. Interestingly, Sophus Lie was the first to introduce continuous symmetries to solve equations and find solutions to differential equations\cite{Olver1986, Bluman2010, Anco1996, Ibragimov1999, Bluman1989, Bluman2002}. Later, the study of these symmetries at the level of solutions became common. Today, these symmetries are more commonly used to find conserved charges and currents, and their role in solving differential equations arising in physics has faded. 

In this section, we will review the idea of Lie for solving differential equations and finding solutions using their symmetries. The symmetries of the equations should be defined first. The symmetry of differential equations includes transformations that keep the form of the equation invariant. Consequently, they transform the solution into another (or the same) solution. The advantage of these transformations is that finding the solutions to differential equations is simpler in canonical coordinates. 

On the other hand, solving the Einstein equations and classifying the solutions, including black hole geometries, is complicated. In higher dimensions, we encounter more involved partial differential equations (see \cite{Emparan-Reall-review} for a review on the classification of black holes in higher dimensions). Near-horizon extremal geometries are also solutions to the Einstein field equations. These geometries emerge in the near-horizon limit of zero-temperature (extremal) black holes. The theorems of \cite{KL-paper-1, KL-paper-2, KL-paper-3} prove a common feature of near-horizon extremal geometries: there is an AdS$_2$ subspace if the parent black hole is asymptotically flat or AdS. The classification of near-horizon extremal geometries is reviewed in \cite{KL-review}. The symmetries and thermodynamics of near-horizon extremal geometries are described by two-dimensional conformal field theories (CFT) which are chiral: they have a single non-trivial Virasoro algebra \cite{Compere}.

A special class of extremal black holes exists for which the horizon areas vanish at the same rate as the temperature, called EVH black holes.  According to theorems in \cite{EVH/CFT} and \cite{Sadeghian:2015}, the near-horizon geometries of extremal black holes with vanishing area include an AdS$_3$ subspace and SO(2,2) symmetry rather than AdS$_2$ and SO(2,1) symmetry for the generic extremal case. In the field theory picture, these geometries are described by 2d CFTs that are not chiral \cite{EVH/CFT}.

The SO(2,2) symmetry not only restricts the functional dependence of a solution but also turns off some components of the metric and matter fields: the metric should include an AdS$_3$ subspace, and the fields of the theory depend only on directions perpendicular to this subspace. Applying the SO(2,2) symmetry to the metric and matter field ansatz and solving the equations of motion, we end with the classification of solutions. This is the procedure carried out in reference \cite{Sadeghian:2014}. 

Provided by the SO(2,2) rigidity theorem in \cite{Sadeghian:2014}, the SO(2,2) symmetry leads to an additional U(1) symmetry for the five-dimensional Einstein-Maxwell-Lambda theory. This additional U(1) symmetry, along with the SO(2,2) symmetry in five dimensions, converts the equations of motion from partial to ordinary differential equations, completing the classification and uniqueness of the solutions. The existence of a scalar field in the theory does not guarantee this additional U(1) symmetry (though it does not negate it either). However, according to the "conjecture" proposed in \cite{Sadeghian:2014}, this additional U(1) symmetry is also a consequence of the SO(2,2) symmetry for theories with scalar fields. 
In the following sections, we use the Lie method to find more solutions (other than the axisymmetric solutions) to the five-dimensional Einstein theories, including Einstein-dilaton, Einstein-Maxwell, and Einstein-dilaton-Maxwell theories with SO(2,2) symmetry, to examine this "conjecture."

%@@@@@@@@@@@@@@@@@@@@@@@@@@@@@@@
\section{5D Einstein-Dilaton Theory}\label{E-D-theory}
%@@@@@@@@@@@@@@@@@@@@@@@@@@@@@@@
In this section, we find solutions with SO(2,2) symmetry for the Einstein-dilaton theory in five dimensions.
The following action describes this theory:
\bea
S=\frac{1}{16 \pi G }\int d^5x\ \sqrt{-g}\ \left(R-2\partial_\mu\Phi\partial^\mu\Phi\right).
\eea
The equations of motion are 
\bea\label{Einstein-dilaton-eq}
R_{\mu\nu}=2\ \partial_\mu\Phi\partial^\mu\Phi,
\eea
\bea\label{phi-eom-1}
\nabla^2\Phi=0,\ \ \ \ 
\eea
For a constant scalar field, the metric functions match the metric solutions of pure Einstein theory classified in \cite{Sadeghian:2014}.

The metric ansatz with SO(2,2) symmetry is
\bea\label{metric-1}
ds^2=e^{2\rho}\ {\rm ds}_{AdS_3}^2+e^{2\sigma}d\theta^2+e^{2\omega}d\psi^2\, ,
\eea
where ${\rm ds}_{AdS_3}^2=-r^2dt^2+\frac{dr^2}{r^2}+r^2d\phi^2$ is the metric of the three-dimensional anti-de Sitter space. In addition, the metric functions $\rho$, $\sigma$, and $\omega$ depend only on $\theta$ and $\psi$ as a result of SO(2,2) symmetry. In \cite{Sadeghian:2014}, the coordinate $\theta$ is chosen so that the dilaton is linear, i.e. $\Phi=\alpha\ \theta$; then equation \eqref{phi-eom-1} transforms into a relation between metric functions: 
\bea\label{f-h-R}
\sigma=\omega+3\rho\,.
\eea
The solutions of this theory, with axisymmetry (along the Killing vector $\partial_\psi$), have been introduced in reference \cite{Sadeghian:2014}. Based on a conjecture addressed in the same reference, it is \textit{assumed} that the additional U(1) symmetry along $\partial_\psi$ exists. Here, we intend to examine the (non-)existence of other solutions that depend on $\psi$ using the Lie method.

Unlike \cite{Sadeghian:2014}, we do not choose the linear dilaton gauge and let the dilaton field depend on the coordinates $\theta$ and $\psi$. The field equations restrict this functional dependence. The dilaton field equation \eqref{phi-eom-1} on the 5-dimensional metric \eqref{metric-1} becomes
\bea\label{phi-eom-2}
\partial_\theta\left(e^{3\rho-\sigma+\omega}\partial_\theta\Phi\right)+\partial_\psi\left(e^{3\rho+\sigma-\omega}\partial_\psi\Phi\right)=0\,.
\eea
For the separability of this equation, we assume that $\Phi(\theta,\psi)=\Phi_1(\theta)+\Phi_2(\psi)$. We define
\bea\label{def-P-T}
T(\theta) \equiv \partial_\theta\Phi_1\,,\quad P(\psi) \equiv \partial_\psi\Phi_2\,.
\eea
We also assume that $T(\theta)$ and $P(\psi)$ are non-zero. The special case $P(\psi)=0$ ($T(\theta)=0$) is studied in \cite{Sadeghian:2014}.
The equation \eqref{phi-eom-2} can be easily solved by considering each parentheses is constant.\footnote{This is not the most general solution to \eqref{phi-eom-2}, $e^{3\rho-\sigma+\omega} T$ can be a function of $\psi$, for e.g. $f_1(\psi)$ and $e^{3\rho+\sigma-\omega} P$ can be a function of $\theta$, as $f_2(\theta)$. For convenience, we fix these degrees of freedom and set $f_1(\psi)=1$ and $f_2(\theta)=1$.} This connects the metric functions $\rho$, $\sigma$ and dilation derivatives, $T$ and $P$, by
\bea
e^{6\rho}=\frac{1}{TP}\,, \quad e^{2\sigma}=\frac{T}{P}\ e^{2\omega}\,,
\eea
Then, the metric ansatz \eqref{metric-1} becomes
\bea\label{metric-2}
ds^2=\frac{1}{\left(T P\right)^{1/3}}\ ds_3^2+\frac{T}{P}\ e^{2\omega} \ d\theta^2+e^{2\omega}d\psi^2\,.
\eea
Given this metric ansatz, the independent components of the Einstein equations \eqref{Einstein-dilaton-eq} are
\bea\label{R-eom-1}
&&S_1)\quad -12\left(\frac{T}{P^2}\right)^{\frac13}e^{2\omega}+\frac{T_{\theta\theta}}{T^2}-2\frac{T_{\theta}^2}{T^3}+\frac{P_{\psi\psi}}{P^2}-2\frac{P_{\psi}^2}{P^3}=0\,,\nnr
&&S_2)\quad 4T+2\frac{\omega_{\theta\theta}}{T}-\frac{T_{\theta\theta}}{T^2}+\frac53\frac{T_{\theta}^2}{T^3}+2\frac{\omega_{\psi\psi}}{P}-2\frac{P_{\psi}h_{\psi}}{P^2}-\frac{P_{\psi\psi}}{P^2}+2\frac{P_{\psi}^2}{P^2}=0\,,\nnr
&&S_3)\quad 8+\frac{\omega_{\theta}}{T}\frac{P_\psi}{P^2}+\frac{T_{\theta}}{T^2}\left(\frac{\omega_{\psi}}{P}-\frac{P_{\psi}}{3P^2}\right)=0\,,\nnr
&&S_4)\quad \frac{\omega_{\theta\theta}}{T}-\frac{T_{\theta}\omega_{\theta}}{T^2}+\frac{\omega_{\psi\psi}}{P}+\frac43\frac{P_{\psi}^2}{P^3}-\frac{P_{\psi\psi}}{P^2}+2P=0\,.
\eea
Based on the Lie method, these equations admit three symmetry generators of the forms
\begin{itemize}
\item[] {\bf I)} $\xi_1=\partial_\theta\ ,$
\item[] {\bf II)} $\xi_2=\partial_\psi\ ,$
\item[] {\bf III)} $\xi_3=\theta\ \partial_\theta+\psi\ \partial_\psi-T\partial_T-P\partial_P-\frac{2}{3}\partial_\omega\ .$
\end{itemize}
We note that the algebra of these generators is given by
\bea\label{algebra}
[\xi_1,\xi_2]=0  \,,\qquad [\xi_1,\xi_3]=\xi_1  \,,\qquad [\xi_2,\xi_3]=\xi_2  \,,
\eea
which is a translation-dilatation algebra. 

In what follows, we investigate the invariant solutions of each generator. As stated bellow \eqref{def-P-T}, we assumed that $T =\partial_\theta \Phi\neq 0$ and $P=\partial_\psi \Phi \neq 0$. Therefore, we already ruled out the invariant solutions under $\xi_1$ and $\xi_2$. Although there exists an invariant solution under $\xi_2$ in the linear dilaton gauge\cite{Sadeghian:2014}, the invariant solutions under $\xi_1$ do not exist even in the linear dilaton gauge. We will come back to the relation between the linear dilaton gauge and our choice of coordinates in \eqref{metric-2} by the end of this section. In what follows, we find the invariant solutions under $\xi_3$.\\
%%%%%%%%%%%%%%%%%%%%%%%%%%%%%%%
The transformations generated by $\xi_3$ are
\bea\label{tr-xi-3}
\theta \to  e^\varepsilon\theta\,,\ \quad \psi\to e^\varepsilon \psi\,,\ \quad P\to e^{-\varepsilon} P\,,\quad T\to e^{-\varepsilon} T\,\quad\ \omega\to \omega - \frac23\varepsilon\,,\ \ \ \ \ \ \ \ 
\eea
and the first prolongation of the infinitesimal generator $\xi_3$ is given by
\bea\label{1st-prolongation}
\xi_3^{(1)}=\theta \partial_\theta+\psi\ \partial_\psi-T \partial_T-P \partial_P-\frac{2}{3}\partial_\omega-\omega_\theta\ \partial_{\omega_\theta}-\omega_\psi \partial_{\omega_\psi}-2 T_\theta \partial_{T_{\theta}}-2 P_\psi \partial_{P_{\psi}}\,. 
\eea
For completeness, we comment on the action of this transformation on $ds^2$ in \eqref{metric-2},
\bea\label{tr-3-metric}
ds^2\to \left(e^{\varepsilon}\right)^{\frac23} ds^2\,.
\eea 
Therefore, this is not an isometry that leaves the metric invariant but acts as a Weyl scaling\footnote{To be precise, the transformation generated by $\xi_3$ is different from the Weyl transformation in which the coordinates are intact, while $\xi_3$ scales the $\theta$ and $\psi$ coordinates as well.}. To find a solution using this symmetry, we write the Einstein equations \eqref{R-eom-1} in terms of the invariant functions of $\xi_3$. The invariant functions under the transformations generated by $\xi_3$ are given by
\bea\label{invs}
&&x\equiv \frac{\psi}{\theta} \,, \quad H \equiv \omega+\frac{2}{3}\ln{(\theta)},\quad \tilde{T}\equiv\theta \cdot T\,,\quad  \tilde{P}\equiv\psi \cdot P\,,\\
&& \tau \equiv \frac{T_\theta}{T^2}\,,\quad \Pi \equiv\frac{P_\psi}{P^2}\,,\quad \Theta \equiv \frac{\omega_\theta}{T}\,,\quad \Psi \equiv \frac{\omega_\psi}{P}\,.
\eea

We write the Einstein equations \eqref{R-eom-1} in terms of the (first-order) invariant functions of $\xi_3$. We observe that the equation $(S_3)$ is an algebraic equation for the functions $\tau$, $\Pi$, $\Theta$ and $\Psi$,
\bea\label{S3}
S_3) \quad 8+\Pi\,\Theta+\tau \left(\Psi-\frac{1}{3}\Pi\right)=0\,.
\eea
Differentiating this equation with respect to $\theta$ gives 
\bea\label{Theta-tau-theta}
\Theta_\theta= \frac{(\Pi-3 \Psi)}{3 \ \Pi}\tau_\theta\,,
\eea
and differentiating equation \eqref{S3} with respect to $\psi$ gives
\bea\label{Psi-Pi-psi}
\Psi_\psi=\left(\frac{\tau-3\ \Theta}{3\ \tau}\right)\Pi_\psi\,.
\eea
We substitute $\Theta_\theta$ in terms of $\tau_\theta$ using \eqref{Theta-tau-theta} and $\Psi_\psi$ in terms of $\Pi_\psi$ by \eqref{Psi-Pi-psi} into the equations $(S_2)$ and $(S_4)$ of \eqref{R-eom-1}, then the Einstein equations \eqref{R-eom-1} simplify to
\bea\label{R-eom-2}
&&S_1) \quad \tau_\theta+ \Pi_\psi -12\left[\frac{T}{P^2}\right]^{\frac13}e^{2\omega}=0\,,\nnr
&&S_2)\quad 
T\left(  \tau^2-6\,\tau\, \Theta -12 \ \right)\,+\left(\frac{\Pi+6\,\Psi}{\Pi}\right) \tau_\theta\,+\left(\frac{ \tau +6\,\Theta}{\tau} \right) \Pi_\psi=0\,,\nnr
&&S_3) \quad 4+\Pi\,\Theta+\frac{\tau}{3} \left(3\ \Psi-\Pi\right)=0\,,\nnr
&&S_4)\quad \left(\frac{3\ \Psi-\Pi}{\Pi}\right)\ \tau_\theta+\left(\frac{3\ \Theta+2\ \tau}{\tau}\right)\Pi_\psi+\left(2\ \Pi^2-3\ \Psi\ \Pi-6\right)\ P=0\,.
\eea
This procedure reduces the order of the differential equations by one. If we further assume $\omega(\theta,\psi)=\omega_1(\theta)+\omega_2(\psi)$ then the separability of the equation $(S_1)$ of \eqref{R-eom-2} leads to 
\bea
&&\Pi_\psi=C_1\,,\label{q1}\\
&&\omega_2-\frac13\ln{P}=C_2\,.\label{q2}
\eea
Differentiating \eqref{q2} with respect to $\psi$ gives 
\bea\label{Pi-Psi}
\Pi=3\ \Psi\,,
\eea
The equation $(S_3)$ of \eqref{R-eom-2} with this constraint forces both $\Pi$ and $\Theta$ to be constant. Then, 
\bea\label{constraint-1}
\Pi_\psi=0\,,
\eea
and 
\bea\label{constraint-2}
\Theta_\theta=0\,.
\eea
This constraint is consistent with equation \eqref{Theta-tau-theta}. In addition, \eqref{constraint-1} and \eqref{Psi-Pi-psi} lead to $\Psi_\psi=0$. Using the constraints \eqref{Pi-Psi} and \eqref{constraint-1} into the equation $(S_4)$ of \eqref{R-eom-2} gives 
\bea
\Pi=-\sqrt{6}\,, \quad \Psi=-\sqrt{\frac{2}{3}}\,.
\eea 
These two equations can be solved for $P(\psi)$ and $\omega_2(\psi)$ 
\bea\label{final-sol-P}
P(\psi)=\frac{1}{\sqrt{6} \ \psi}\,, \quad \omega_2=-\frac{1}{3}\ln{\psi}+C_3\,.
\eea
Taking into account these results with the equation $(S_3)$ of \eqref{R-eom-2}, we have 
\bea\label{Theta-sol}
\Theta=\frac{4}{\sqrt{6}}\,.
\eea
Finally, the simplified version of $(S_1)$ and $(S_2)$ are
\bea\label{eq-tau-theta}
&& S_1)\quad \tau_\theta-12\ e^{2\left[\omega_1+\frac16 \ln{T}\right]}=0\,,\nnr 
&& S_2)\quad \tau_\theta+\left[\frac{\tau^2}{3}-\frac{8}{\sqrt{6}}\tau-4\right] \ T=0\,.
\eea
We remind that the function $\tau(\theta)$ is related to $T(\theta)$ by $\tau=T_\theta/T^2$. We observe that the transformation
\bea\label{tr-theta-tilde}
T=e^{-6 u(\tilde{\theta})}\,;\quad \frac{\partial \tilde{\theta}}{\partial \theta}=T(\theta)\,,% \quad d\tilde{\theta}= T(\theta)\ d\theta\,,
\eea
simplifies the equations \eqref{eq-tau-theta} to 
\bea\label{rho-tilde-theta}
&&S_1)\quad u_{\tilde{\theta} \tilde{\theta}}+2\ e^{2\omega_1+4\rho}=0\,, \nnr
&&S_2)\quad u_{\tilde{\theta} \tilde{\theta}}-2\ u_{\tilde{\theta}}^2-\frac{8}{\sqrt{6}}u_{\tilde{\theta}}+\frac23=0\,.
\eea
The solution to these equations is given by
\bea\label{final-u-sol}
u(\tilde{\theta})=-\frac12\ln{\left(\cosh{\left[k(\tilde{\theta}+C_4)\right]}\right)}-\frac{2}{\sqrt{6}}\ \tilde{\theta}\,,
\eea
here, $k=2$. This reminds us of the metric function found in \cite{Sadeghian:2014}. 
If we rewrite the relation \eqref{Theta-sol} in terms of $\tilde{\theta}$, we have
\bea\label{h-tilde-theta}
\omega_{\tilde{\theta}}=\frac{\omega_{\theta}}{T}=\frac{4}{\sqrt{6}}\,,
\eea
which can be solved by
\bea\label{final-h1-sol}
\omega_1(\tilde{\theta})=\frac{4}{\sqrt{6}}\ \tilde{\theta}\,.
\eea
Looking back at the transformation \eqref{tr-theta-tilde}, which relates $\tilde{\theta}$ to $\theta$, we can better understand how this changes $\theta$ to the linear dilaton gauge
\bea
T=\partial_\theta \Phi=\frac{\partial \tilde{\theta}}{\partial \theta}\ \partial_{\tilde{\theta}}\Phi=T\  \partial_{\tilde{\theta}}\Phi\,.
\eea
Then $\partial_{\tilde{\theta}}\Phi=1$, which justifies the linear dilaton gauge. 

The final step to relate our new solution and the axisymmetric solution in \cite{Sadeghian:2014} in the linear dilaton gauge is the following transformation from $\psi$ to $\tilde{\psi}$
\bea
{\psi}=\frac{{ \tilde{\psi}}^2}{2}\,,
\eea
which takes the metric \eqref{metric-2} to a $\tilde{\psi}$-symmetric solution 
\bea\label{final-metric-2}
ds^2=e^{\frac23\ln{\tilde{\psi}}}\left[A\ e^{2u}ds_{AdS_3}^2+ A\ e^{6u+2\omega_1}\ d\tilde{\theta}^2+e^{2\omega_1}\ d\tilde{\psi}^2\right]\,.
\eea
Here, $A$ is a numeric constant and the metric functions $\omega_1(\theta)$ and $u(\theta)$ are given by \eqref{final-u-sol} and \eqref{h-tilde-theta}. In terms of $\tilde{\theta}$ and $\tilde{\psi}$ coordinates, the dilaton field solution to the Einstein-dilaton theory takes the following form
\bea\label{final-dilaton-2}
\Phi(\tilde{\theta},\tilde{\psi})=\tilde{\theta}+\sqrt{\frac{2}{3}}\ln{\tilde{\psi}}+\Phi_0\,.
\eea
The metric in brackets can be compared to the $\psi$-symmetric solution found in \cite{Sadeghian:2014} in the linear dilaton gauge, 
\bea
ds^2_0=e^{2R(\tilde{\theta})} ds_{AdS_3}^2+e^{ 6R(\tilde{\theta})+2h({\tilde{\theta}}) }d\tilde{\theta}^2+e^{ 2 h(\tilde{\theta}) }d\tilde{\psi}^2\,.
\eea
This shows that, up to some trivial transformations by some constant factors, the new solution we find here is scaled with respect to the $\psi$-symmetric solution in \cite{Sadeghian:2014}. The scale factor is $\Omega^2=e^{\frac23\ln{\tilde{\psi}}}$.

Studying the conformal Killing equations for this solution \eqref{final-metric-2}, whose metric functions are given by \eqref{final-sol-P}, \eqref{final-u-sol}, \eqref{final-h1-sol}, we find that it has seven conformal Killing vectors, including $\partial_\psi$ and obviously the generators of SO(2,2) symmetry.

We note that the metric \eqref{final-metric-2} has curvature singularity at $\psi=0$ since the Kretschmann scalar for the metric \eqref{final-metric-2} has the following form
\bea
\mathcal{K}^2=\frac{k_1(\theta)}{A^2\psi^{4/3}}+\frac{k_2(\theta)}{A \psi^{10/3}}+\frac{k_3(\theta)}{\psi^{16/3}}\,.
\eea
Here, $k_1(\theta)$, $k_2(\theta)$, $k_3(\theta)$ are regular functions. $(\nabla \Phi)^2$ is also singular at $\psi=0$.

%@@@@@@@@@@@@@@@@@@@@@@@@@@@@@@@
\section{5D Einstein-Maxwell Theory}\label{E-M-theory}
%@@@@@@@@@@@@@@@@@@@@@@@@@@@@@@@
The action of Einstein-Maxwell theory is 
\bea
S=\frac{1}{16 \pi G\ }\int d^5x\ \sqrt{-g}\ \left(R-F_{\mu\nu}F^{\mu\nu}\right)\,.
\eea
Subsequently, the equations of motion are
\bea
&&R_{\mu\nu}=2F_{\mu\lambda}F_\nu^\lambda-\frac{1}{3}g_{\mu\nu}F_{\sigma\lambda}F^{\sigma\lambda},\nnr
&&\nabla_\mu F^{\mu\nu}=0.\ \ \ \ \ 
\eea
According to the theorem in \cite{Sadeghian:2014}, the SO(2,2) symmetry leads to additional axisymmetry. However, we recheck this result with the Lie symmetry method and examine the symmetries of the equations of motion for this theory.
We use the metric ansatz in \eqref{metric-1}, which is invariant under the SO(2,2) group. \footnote{We consider a different metric ansatz from the one proposed in \cite{Sadeghian:2014}. There, Theorem \cite {Sadeghian:2014} restricts the metric functions to depend only on $\theta$. Here, we do not assume axisymmetry to derive it from the equations of motion.} and we emphasise that the metric functions $\rho$, $\sigma$, and $\omega$ depend on both $\theta$ and $\psi$.

Applying SO(2,2) symmetry to the Maxwell field restricts the Maxwell field components:  the only non-zero component of the Maxwell field is \( F_{\theta\psi} \).  The Maxwell equation expresses this component in terms of metric functions by
\bea
F_{\theta\psi}=p\ e^{\sigma+\omega-3\rho}\,.
\eea
By substituting it into the Einstein equations, the following differential equations are obtained
\bea
&&\rho_{\theta\psi}+\rho_\theta \rho_\psi-\rho_\theta \sigma_\psi-\rho_\psi \omega_\theta=0,\nnr
&&\rho_{\theta \theta}+\rho_\theta\left(\omega_\theta+3\,\rho_\theta-\sigma_\theta\right)+2\,{{\rm e}^
{2\,\sigma-2\,\rho}}+\left(\rho_{\psi \psi} -\rho_\psi \omega_\psi+3\, \rho_\psi^2+\rho_\psi \sigma_\psi\right) {\rm e}^{2\,\sigma-
2\,\omega}=\frac23\,{p}^{2}{{\rm e}^{2\,\sigma-6\,\rho}}\,,\nnr
&&\rho_{\theta \theta}+\frac13 \omega_{\theta \theta}+\rho_\theta\left(\rho_\theta-\sigma_\theta\right)+\frac13 \omega_\theta \left(\omega_\theta -\sigma_\theta  \right)+\frac13\left(\sigma_{\psi \psi} +\sigma_\psi\left[\sigma_\psi- \omega_\psi+3\, \rho_\psi\right]\right) {\rm e}^{2\,\sigma-
2\,\omega}\nnr
&&\hspace{12cm}=-\frac49\,{p}^{2}{{\rm e}^{2\,\sigma-6\,\rho}}\,,\nnr
&&\frac13\left(\omega_{\theta \theta}+\omega_\theta\left[\omega_\theta-\sigma_\theta+3\rho_\theta\right]\right)+\left(\rho_{\psi\psi}+\frac13\sigma_{\psi\psi}+\rho_\psi^2-\rho_\psi \omega_\psi+\frac13\sigma_\psi^2-\frac13 \omega_\psi \sigma_\psi\right)e^{2\sigma-2\omega}\nnr
&&\hspace{12cm}=-\frac49p^2 e^{2\sigma-6\rho}\,.\nnr
\eea
The symmetry generators of these equations are given by
\bea
K=F_1(\theta)\ \partial_\theta+F_2 (\psi)\  \partial_\psi-F_{1,\theta} \ \partial_\sigma-F_{2,\psi} \ \partial_\omega\, . 
\eea
It is easy to see that this generator keeps $ds^2$ invariant. Thus, it acts as an isometry. A special case of these transformations is $F_2 (\psi)=1$ (or $F_1 (\theta)=1$), and the invariant solutions under this generator are axisymmetric as predicted by Theorem 2 in \cite{Sadeghian:2014}. Therefore, this symmetry generator (K) cannot be used to construct new solutions other than those trivially related to the solutions presented in \cite{Sadeghian:2014}.

%@@@@@@@@@@@@@@@@@@@@@@@@@@@@@@@
\section{5D Einstein-Maxwell-Dilaton Theory}\label{E-M-D-theory}
%@@@@@@@@@@@@@@@@@@@@@@@@@@@@@@@
This theory does not meet the conditions of the theorems presented in \cite{Sadeghian:2014} because it has a dilaton field. Therefore, this theory may have a solution that lacks axisymmetry. The following action describes the Einstein-Maxwell-dilaton theory
\bea
S=\frac{1}{16\ \pi\ G\ }\int d^5x\ \sqrt{-g}\ \left(R-2\partial_\mu\Phi\partial^\mu\Phi-F_{\mu\nu}F^{\mu\nu}\right),
\eea
and the equations of motion are
\bea
&&R_{\mu\nu}=2\ \partial_\mu\Phi\partial^\mu\Phi+2F_{\mu\lambda}F_\nu^\lambda-\frac{1}{3}g_{\mu\nu}F_{\sigma\lambda}F^{\sigma\lambda}\,,\nnr
&&\nabla_\mu F^{\mu\nu}=0\,,\nnr
&&\nabla^2 \Phi=0\,.\  
\eea
The discussion of the scalar field is the same as in section \ref{E-D-theory}. So, to simplify the equations, we use the linear dilaton gauge $\Phi=\alpha \ \theta$. Applying the SO(2,2) symmetry to the Maxwell field, we find the non-zero component is $F_{\theta\psi}$, and the Maxwell equation implies $F_{\theta\psi} = p \, e^{2\omega}$. Substituting this result and the metric ansatz \eqref{metric-1} (with the constraint \eqref{f-h-R}) into the equations of motion, we obtain the following differential equations
\bea\label{EMD-diff-eq}
&&\hspace{-1cm}\rho_{\theta\theta}+2\ e^{2\omega+4\rho}+\left(\rho_{\psi\psi}+6\rho_\psi^2\right)e^{6\rho}=\frac{2}{3}p^2e^{2\omega},\ \ \ \ \ \nnr
&&\hspace{-1cm}\rho_{\theta\theta}+\frac{1}{3}\omega_{\theta\theta}-2\rho_\theta^2-2\ \omega_\theta \rho_\theta+\left[\rho_{\psi\psi}+\frac{1}{3}\omega_{\psi\psi}+6\rho_\psi^2+2\omega_\psi \rho_\psi\right]e^{6\rho}=-\frac{4}{9}p^2e^{2\omega}-\frac{2}{3}\alpha^2,\nnr
&&\hspace{-1cm}\rho_{\theta\psi}-2\rho_\theta \rho_\psi-\omega_\theta \rho_\psi-\omega_\psi \rho_\theta=0,\nnr
&&\hspace{-1cm}\frac{1}{3}\omega_{\theta\theta}+\left(2\rho_{\psi\psi}+\frac{1}{3}\omega_{\psi\psi}+4\rho_\psi^2\right)e^{6\rho}=-\frac{4}{9}p^2e^{2\omega}\,.\ 
\eea
The symmetry generators of these equations are $\zeta_1=\partial_\theta,\ \zeta_2=\partial_\psi$. The axisymmetric solutions that are invariant under $\zeta_2$ are studied in \cite{Sadeghian:2014}. Regarding the invariant solutions under $\zeta_1$, we drop the $\theta$-dependence in \eqref{EMD-diff-eq} and have to solve the following equations of motion 
\bea
&&\left(\rho_{\psi \psi}+6\ \rho_\psi^2\right)\ e^{6\rho}+2 \ e^{2\omega+4\rho}-\frac23\ p^2 \ e^{2\omega}=0\,, \nnr
&& \left(\rho_{\psi\psi}+\frac13 \omega_{\psi\psi}+2\omega_\psi \rho_\psi+6\rho_\psi^2\right) e^{6\rho}+\frac49\ p^2\ e^{2\omega}+\frac23 \ \alpha^2=0\,,\nnr
&&\left(2\rho_{\psi\psi}+\frac13\omega_{\psi\psi}+4 \rho_\psi^2\right)e^{6\rho}+\frac49\ p^2\ e^{2\omega}=0\,.
\eea
These equations are inconsistent and do not yield any $\theta$-symmetric solutions. 

%@@@@@@@@@@@@@@@@@@@@@@@@@@@@@@@
\section{Discussion and outlook}
%@@@@@@@@@@@@@@@@@@@@@@@@@@@@@@@
In this work, we find a novel SO(2,2)-symmetric solution to five-dimensional Einstein-dilaton theory that lies beyond the class of solutions classified in \cite{Sadeghian:2014}. It is constructed by applying the Lie algorithm to the equations of motion. The symmetries of the equations of motion of the five-dimensional Einstein-dilaton theory include two U(1) symmetries along the $\xi_1=\partial_\theta$ and $\xi_2=\partial_\psi$ directions and a symmetry along the generator $\xi_3=\theta\ \partial_\theta+\psi\ \partial_\psi-T\partial_T-P\partial_P-\frac{2}{3}\partial_\omega$. This symmetry generator acts on both the space-time coordinates and the metric functions and leaves the Einstein equations of motion invariant. We study the action of $\xi_3$ on the geometry and find that $\xi_3$ acts on $ds^2$ as a Weyl scaling, $ds^2\to e^{\frac23\varepsilon} ds^2$. Therefore, the weight of $ds^2$ under this scaling is $\Delta=\frac23$. This weight arises from the space-time dimension ($d=5$) and the dimension of the AdS$_3$ subspace. 

The invariant solutions under $\xi_1$ ($\xi_2$) are discussed in \cite{Sadeghian:2014}. Assuming separability of the equations of motion and using the $\xi_3$ invariant functions, we reduce the order of the differential equations from 2 to 1 and find the solution. Other solutions to the equations of motion may exist that are not invariant under these symmetries or separable. We hope to address it in the future.

It would be interesting to find a black hole/string/brane solution to the 5d Einstein-dilaton theory whose near-horizon EVH geometry is described by the new solution we found here in \eqref{final-metric-2}, \eqref{final-sol-P}, \eqref{final-u-sol}, \eqref{final-h1-sol}. 

Related to the symmetries of the geometry, there are conserved charges\cite{Wald-1, Wald-2, Wald-3}. Moreover, one can associate conserved charges to the Lie symmetries of differential equations\cite{Olver1986}. We concern ourselves with the relation between these two setups and study them for the novel solution and also in a general case.
Applying dimensional reduction on the 2-dimensional space parameterized by $\theta, \psi$, we end up with an asymptotically AdS$_3$ space. Understanding the dual CFT  is another interesting question we postpone to the future\cite{Compere:2014bia}.

For the five-dimensional Einstein-Maxwell theory, we find no new solutions using the Lie symmetry method. This is in accordance with the theorems of \cite{Sadeghian:2014} on the axisymmetry of the solutions. By a new solution, we mean a non-trivial solution that was not obtained by redefining each coordinate with a function of itself. Although no such theorem exists for the Einstein-Maxwell-dilaton theory, we could not find new solutions invariant under the symmetries of the equations of motion.

\acknowledgments
The author thanks M. M. Sheikh-Jabbari, H.R. Afshar, and S. Aghapour for the comments and discussions.

%\paragraph{Note added.} This is also a good position for notes added after the paper has been written.

% Bibliography

%% [A] Recommended: using JHEP.bst file
%% \bibliographystyle{JHEP}
%% \bibliography{biblio.bib}

%% or
%% [B] Manual formatting (see below)
%% (i) We suggest to always provide author, title and journal data or doi:
%% in short all the informations that clearly identify a document.
%% (ii) please avoid comments such as "For a review'', "For some examples",
%% "and references therein" or move them in the text. In general, please leave only references in the bibliography and move all
%% accessory text in footnotes.
%% (iii) Also, please have only one work for each \bibitem.

\end{document}